\documentclass{aastex}
 
\def\myputfigure#1#2#3#4%
{\hskip0.03\textwidth
\makebox[0pt]{\hskip#2in
\includegraphics[width=#3\textwidth]{#1}}\vskip#4pt\hfill}
 
\usepackage{emulateapj5, graphicx}

\begin{document}

\title{Understanding Cluster Gas Evolution and Fine-Scale CMB Anisotropy 
with Deep Sunyaev-Zel'dovich Effect Surveys}
\shorttitle{Gas Evolution with SZE Surveys}
\author{Gilbert P. Holder and John E. Carlstrom}
\affil{Department of Astronomy and Astrophysics, 
University of Chicago,
Chicago, IL 60637}
\email{holder@oddjob.uchicago.edu,jc@hyde.uchicago.edu}

\begin{abstract}

We investigate the impact of gas
evolution on the expected yields from deep Sunyaev-Zel'dovich (SZ)
effect surveys as well as on the expected SZ effect contribution
to fine scale anisotropy in the Cosmic Microwave Background.  The
approximate yields from SZ effect surveys are remarkably
insensitive to gas evolution, even though the observable
properties of the resulting clusters can be markedly different.
The CMB angular power spectrum at high multipoles ($\ell \ga
2000,\ \la 5'$) due to the SZ effect from clusters is quite
sensitive to gas evolution. We show that moderate resolution
($\leq 1'$) SZ effect imaging of clusters found in deep SZ effect
surveys should allow a good understanding of gas evolution in
galaxy clusters, independent of the details of the nature of the
gas evolution. Such an understanding will be necessary before
precise cosmological constraints can be set from yields of large
cluster surveys.

\end{abstract}

\keywords{cosmic microwave background --- cosmology: theory ---
large-scale structure of universe --- cosmological parameters}

\section{Introduction}

The Sunyaev-Zel'dovich (SZ) effect has recently become a valuable
observational tool \citep{birkinshaw99}. In the past few years,
observational programs have shifted from attempting to detect the
effect to using the SZ effect to map out massive clusters of galaxies,
study the intra-cluster medium (ICM), and constrain cosmological 
parameters \citep{carlstrom00}.

The thermal SZ effect is a distortion of the cosmic microwave
background (CMB) spectrum caused by hot gas along the line of sight to
the surface of last scattering. The cool
CMB photons undergo Compton scattering on the hot electrons,
gaining on average a small amount of energy in the process,
creating
an intensity decrement at low frequencies ($\nu \la 218$GHz)
and an increment at high frequencies. 

Current instruments are now regularly detecting and imaging 
clusters at high 
signal-to-noise, and the next generation of instruments should be 
capable of mapping fairly large portions of the sky as a means of 
finding clusters of galaxies. 
Several works (e.g., Bartlett 2000; Holder {\em et al.} 2000;
Barbosa {\em et al.} 1996; Kneissl {\em et al.} 2001) 
have predicted the number of clusters that could be
expected in future SZ surveys and their angular power spectrum. 
The survey yields are quite impressive; the next generation of SZ instruments
should be able to detect several clusters per day.

In this work, we take a slightly different perspective.  Rather than
asking how many clusters an SZ survey will find, we
instead focus on which clusters we would like to find and what we can learn 
from these objects.
We use simple models for the gas distribution of galaxy
clusters to investigate the impact of non-gravitational heating on 
observable cluster properties, focusing
on the SZ effect. 
Much work has gone into using the X-ray emission from 
nearby groups and low-mass clusters to constrain feedback from galaxy 
formation and non-gravitational heating 
(e.g., Bower {\em et al.} 2000; Cavaliere, Menci, and Tozzi 1999;
Balogh, Babul, and Patton 1999) or alternatively the 
preferred cooling of low-entropy gas \citep{bryan00}.
The effects of feedback or cooling 
should be even more important at higher redshift and 
for low masses, where the SZ effect will be an
important probe. 

In \S\ref{sec:models} we outline our models and the important SZ observables.
We show the expected SZ properties of clusters
in \S\ref{sec:props} and angular power spectra for these models in
\S\ref{sec:cl}.  
Finally, in \S\ref{sec:disc}, we discuss our results and implications
for planned SZ surveys.  

\section{SZ Effect From Galaxy Clusters}
\label{sec:models}

X-ray observations indicate that the cluster gas is approximately isothermal
\citep{irwin00}, and that the gas density profile can be approximated by
a ``beta model'' \citep{mohr99}:
\begin{equation}
\rho_g(r) = { \rho_{g,\circ} \over [1+(r/r_c)^2]^{3\beta/2}} \quad,
\end{equation}
where $\rho_{g,\circ}$ is the central gas density and $r_c$ is a core
radius. 
We choose to fix $\beta=2/3$, with the knowledge that this will give a
reasonable SZ profile even if the true SZ best-fit value of $\beta$ for 
clusters is slightly higher \citep{reese00,grego00b}. 
With this choice of $\beta$, the mass is divergent, and
must be truncated at some radius. We set the truncation radius to be the
virial radius, derived from the spherical collapse model, as outlined below.

For this cluster model, the central decrement of the cluster can be written
as
\begin{equation}
{\Delta T_\circ \over T_{CMB}} = f(\nu) \ y = 
2 f(\nu) {kT_e \over m_e c^2} \sigma_T
  n_\circ r_c \arctan{({R_v \over r_c})} \quad ,
\label{eqn:dec}
\end{equation}
where $y$ is the Compton $y$ parameter, 
$R_v$ is the virial radius, $T_e$ is the electron temperature,
$n_\circ$ is the central electron number density, $\sigma_T$ is the
Thomson cross-section, and $f(\nu)$ is a dimensionless frequency factor
(e.g., see Birkinshaw 1999 \nocite{birkinshaw99}),
which we will take to be -2 (i.e., the Rayleigh-Jeans limit).
The central
decrement is redshift-independent for a given set of cluster properties,
allowing a distance-independent probe of cluster properties. 

An observable which is more readily interpreted is the integrated SZ flux
density from clusters. Assuming that the cluster is approximately isothermal
and noting that an integral over solid angle can be written as 
$d\Omega= dA/d_A^2$, we can write
\begin{equation}
S = S_\circ \int d\Omega {\Delta T \over T_{CMB}} 
  =  S_\circ f(\nu)  {kT_e \over m_e c^2} \sigma_T {N_e \over d_A^2(z)}
\quad ,
\label{eqn:flux}
\end{equation}
where $N_e$ is the total number of electrons, $d_A$ is the angular diameter
distance and $S_\circ$ gives the conversion from temperature to flux density. 
At a frequency of 30 GHz, the conversion between flux and 
$Y\equiv \int y d\Omega$ is approximately $S = 12.7 (Y/arcm^2)$ Jy.
The mass distribution does not
matter in this case, as long as the integration over solid angle extends
to the cluster virial radius and
the cluster is roughly isothermal.  

The simple relation between the total
flux density and the total mass and
electron temperature make the SZ a particularly valuable tool for searching
for clusters. Cluster detections are largely decoupled from the details
of the structure of clusters, allowing an unbiased sample for studies of both 
cosmology and cluster properties at any redshift. 

\subsection{Mass-Temperature Relation}

In connecting with SZ observables, a
relation is needed between mass and temperature. Such a relation can be
derived from the spherical collapse model \citep{lahav91} or can be
found from numerical simulations. We have chosen the hybrid approach of
Bryan and Norman (1998),
\nocite{bryan98} where the functional form is taken from the spherical 
collapse model and the normalization is found from simulations. 

Using a notation where $H(z) \equiv 100h \, E(z) ~{\rm km~s^{-1}~Mpc^{-1}}$, we use
\begin{equation}
T_{gas} = 1.4 \times 10^7 (\Delta(z) E(z)^2)^{1 \over 3}
M_{15}^{2 \over 3} \, \, {\rm K} \quad ,
\label{eqn:mt}
\end{equation}
with $M_{15}$ the virial mass in units of $10^{15} h^{-1} M_\odot$,
$\Delta(z)$ is the  mean cluster density relative to the critical
density at that redshift and the virial radius is defined as the radius that 
encloses this mean density.

\subsection{Evolution of Cluster Gas}

The SZ observables will be sensitive to the density profile of the ICM,
so it is important to model the ICM distribution both as a function of
mass and as a function of redshift. The simplest evolution is
self-similar evolution, where the central density scales with the
background density and sizes will be proportional to the cosmic scale factor. 
In the absence of physics other than gravity, we would expect something like 
this sort of evolution, since there is no preferred scale. 

It is well-known(e.g., Kaiser 1991; Evrard and Henry 1991; Ponman, Cannon
and Navarro 1999)
that X-ray observations suggest that gas evolution is not self-similar. 
If reionization added significant entropy to the gas or if there has been
substantial feedback into the ICM from galaxy formation, the excess entropy
would shift the ICM to a higher adiabat, leading to a final state of the
cluster that is less compact. In such a situation, the outer parts of the
cluster might be expected to be largely unchanged, but the details of the
cluster profile might depend on the details of the
entropy injection (or preferential removal of low entropy gas).

We will assume that the outer parts of clusters are unaffected by
feedback or preheating, and approximate the dark
matter density profile as a singular isothermal sphere,
while the gas density has a well-defined core that is set by the
minimum entropy of the gas.  For self-similar evolution, this central
minimum entropy is set by gravitational processes, while preheated models
will have extra entropy injected by other processes, which may or may
not be larger than the entropy from gravitational heating.

As a simple model, we adopt the notion of an entropy floor \citep{ponman99},
where feedback has provided a uniform amount of entropy, and
gravitational heating has provided an amount of entropy that increases with
increasing mass. This is a slight modification of the constant entropy core
model of Evrard and Henry (1991)\nocite{evrard91}. With an
entropy floor rather than a fixed central entropy, high-mass clusters can
have a central entropy which is significantly higher than the entropy
floor, more in line both with physical intuition and observations.

We define a measure of the entropy as
\begin{equation}
s \equiv {T \over n_e^{2/3}}   \qquad  .
\label{eqn:entropy}
\end{equation}
and take the central entropy to be the entropy expected from self-similar
evolution plus a constant entropy floor. 

The effect of the central entropy floor is to set a maximum central density 
for a given temperature. For high-mass clusters, the central entropy is set
by the gravitational heating;
for low-mass clusters, the entropy from gravitational heating is 
relatively small and the central entropy is determined by the
entropy floor, leading to a temperature-dependent central density. 

\subsubsection{Self-Similar Evolution}

For self-similar evolution ($s_{floor}=0$), 
the core radius should be a fixed fraction of the virial radius. 
We choose this ratio $\mathcal{R} \equiv R_v/r_c=10$, in rough agreement
with observations of clusters \citep{mohr99}. 
Under the assumption that the gas density at the virial 
radius is equal to the global baryon fraction,
$f_B\equiv\Omega_B/\Omega_m$, times the 
mass density at the virial 
radius, we solve for the central density:
\begin{equation}
n_{\circ;ss} = {f_B \over 3 \mu_e m_H} \Delta(z) \rho_{crit}(z)
	(1 + \mathcal{R}^2) \qquad .
\label{eqn:nss}
\end{equation}

In general, the global baryon fraction is not equal to the cluster gas mass
fraction. Simulations without feedback from galaxy formation
typically find values for the cluster gas mass fraction that
are only slightly lower than the input global baryon fraction \citep{evrard97}.
As we outline below, it is possible
for the cluster gas mass fraction to deviate strongly from the global baryon
fraction if non-gravitational heating is introduced.

The entropy expected from self-similarity for a cluster of a given mass
is easily obtained from  equation \ref{eqn:entropy} using equations
\ref{eqn:mt} and \ref{eqn:nss}, and can be seen to
simply scale with the virial temperature. 

The central SZ decrement assuming self-similar evolution and 
solar abundances of hydrogen and helium ($\mu_e=1.14$), from  
equation \ref{eqn:dec}, is
\begin{equation}
\Delta T_\circ = 295 \ \mu K \ 
f_B h \ {1+\mathcal{R}^2 \over \mathcal{R}} \arctan{\mathcal{R}} \ M_{15} 
\Delta_{178} E(z)^2 \qquad  .
\label{eqn:mod_dec}
\end{equation}
The total flux density,
assuming an observing frequency of 30 GHz, 
can be calculated from equation \ref{eqn:flux}:
\begin{eqnarray}
S = 146 \ {\rm mJy} \ f_B h \ M_{15}^{5/3} 
( \Delta _{178} E(z)^2 )^{1/3}   \nonumber
\\
{ (1+\mathcal{R}^2)(\mathcal{R} -\tan^{-1} \mathcal{R}) \over \mathcal{R}^3} 
\Bigl({ d_A(z) \over 1000 h^{-1} {\rm Mpc}} \Bigr)^{-2} 
\quad ,
\label{eqn:mod_flux}
\end{eqnarray}
where $\Delta_{178}\equiv \Delta/178$. In a universe with $\Omega_m=1$ the
spherical collapse model predicts $\Delta=178$, with smaller values for 
low-density universes.

\subsubsection{``Preheated'' Models}

For a given temperature, the central density is fixed by the central
entropy, and we can use equation \ref{eqn:nss} to solve for the
appropriate value of the core-to-virial ratio $\mathcal{R}$.  For the
self-similar case the central entropy simply scales with $T$ and 
$\mathcal{R}=constant$.  A non-zero entropy floor breaks
the self-similarity and $\mathcal{R}$ becomes a function of temperature.
We assume that the entropy from gravitational heating is equal to that
expected in the self-similar case and add entropy from 
non-gravitational heating.
We can still use equations \ref{eqn:mod_dec} and \ref{eqn:mod_flux}, but
we must now solve for $\mathcal{R}$ as a function of mass. 
For low-mass clusters, this leads to significantly different evolution and 
appearance, as well as a depressed cluster gas mass fraction. 
As an example, for $\Omega_m=1$, the self-similar case would
predict that the central decrement would scale as $M(1+z)^3$, whereas
the entropy floor would predict a scaling as $M^{3/2}(1+z)^{9/4}$.
This relation is different in both its mass scaling
and its redshift dependence. 
In this scenario, low-mass high-redshift 
clusters will be significantly less compact than in a self-similar picture. 
This would have observable consequences, both in the properties of 
high-redshift clusters and in their signature in CMB experiments
(see \S 4).  

For clusters well below $\sim 10^{14}h^{-1} M_\odot$ the 
core radius can become significantly larger than the virial radius
for high values of the entropy floor. In such cases, the gas
distribution is still truncated at the virial radius in our models. 
This is simply
indicating that the entropy of the central gas is sufficiently high that the
central density is not significantly higher than the gas density near the 
virial radius. In this regime, the accretion process could be seriously 
affected by the entropy floor and it is not clear that our simple model is 
applicable.

In this work, we assume a relatively high value for the
entropy floor of $s_{floor}= 200~{\rm keV~cm^2}$ as an extreme case.
Current estimates of the entropy floor \citep{ponman99} are roughly
$s_{floor}=100~{\rm keV~cm^{2}}$.

\subsection{Cosmological Evolution of Cluster Abundances}

We modeled the cosmological evolution of cluster abundances with the
Press-Schechter prescription \citep{press74,bond91}, which gives
the comoving number density as a function of both mass and redshift. 
We followed the procedure outlined in Holder et al. (2000), with two minor 
modifications.  The power spectrum was computed using the fitting functions of 
\cite{eisenstein99a}, and the redshift evolution of the 
power spectrum was evaluated numerically from linear theory 
(e.g., Peebles 1980).

Given the comoving number density, it is straightforward to obtain the
number of clusters per steradian above some mass threshold $M_{lim}$ as
a simple integral over the comoving number density.
To estimate the angular power spectrum of the thermal SZ, we
follow Cole and Kaiser (1988)\nocite{cole88}, 
and assume that only Poisson contributions to the angular 
power spectrum are important.

For this work, we assume that 
$\Omega_m=0.3,\ \Omega_\Lambda=0.7, \ \sigma_8=1$ and $h=0.65$. 
The expected counts are very sensitive to cosmological parameters, as
many authors have found.

\section{Probing High-Redshift Clusters}
\label{sec:props}

The effects of preheating on cluster structure can be seen in Figure 
\ref{fig:dt_rc}.  The largest effects of preheating are seen at low
masses and high redshift. These models have been designed to agree with 
observed properties of X-ray emitting clusters and groups at low redshift,
so the true test of these models will be how they fare at high redshift.
In particular, important information can be learned about cluster-to-cluster
variations in the amount of non-gravitational heating and also in the
redshift evolution of the amount of preheating in clusters.

\myputfigure{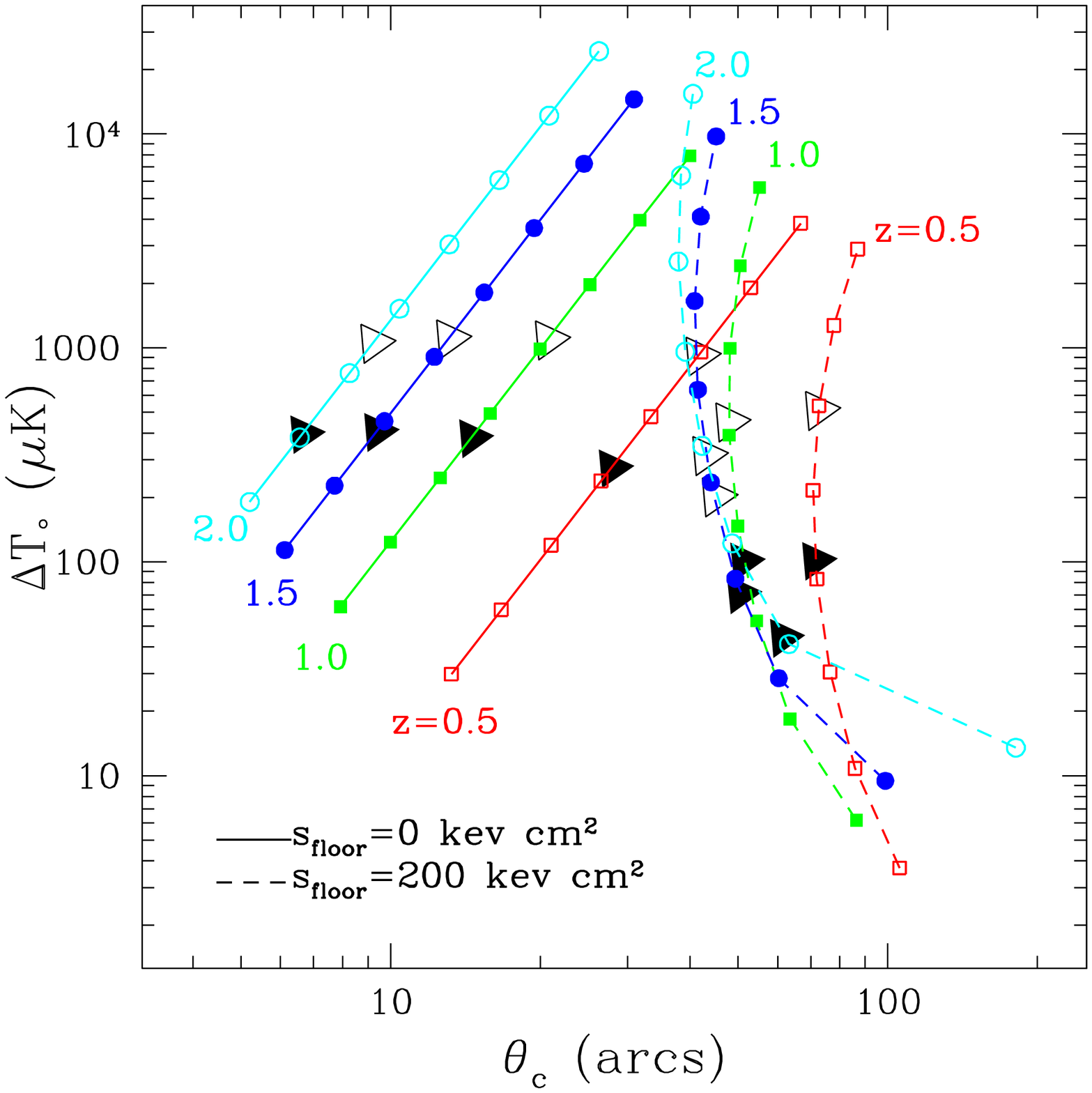}{2.8}{0.45}{-15}
\figcaption{Central decrement and core radius for several masses
at several redshifts ($z=0.5,1,1.5,2$) for  a preheated model
($s_{floor}=200 {\rm keV\, cm^2}$) and self-similar evolution
($s_{floor} =0$).  Mass is marked by open squares along each line and 
increases along each line from bottom to top in
factors of 2, from $0.5-64\times10^{14}h^{-1}M_{\odot}$. Lines of
constant redshift are marked with similar symbols.
Solid (open) triangles mark the limiting mass at which $dN/dz>1$(0.01) 
per square degree
per unit redshift for a $\Lambda CDM$ cosmology.
\label{fig:dt_rc}}

\vskip 0.1in

In Figure 1, we have marked the limiting survey
mass at each redshift which is required to obtain a surface density per 
redshift bin 
of either 1 or 0.01 clusters
per square degree per unit redshift.
In order to get a large sample of clusters at high 
redshift, the survey mass limit must be relatively low. 

Typical expected mass limits for SZ surveys are shown in 
Figure \ref{fig:mlims}.  PLANCK is expected to be able to survey most of 
the sky, at a resolution of $\sim 5'$ down to a level of $\sim 10\mu K$. 
Deep ground-based
surveys should be able to reach a comparable temperature uncertainty on a few
tens of square degrees at a resolution of $\sim 1'$.

The SZ is well
suited for surveying, with the nice feature that the survey mass limit 
should be largely decoupled from issues of preheating, as illustrated
in Figure \ref{fig:mlims} by
the small difference in the mass limits between the self-similar model and 
the fairly extreme preheated model.

\myputfigure{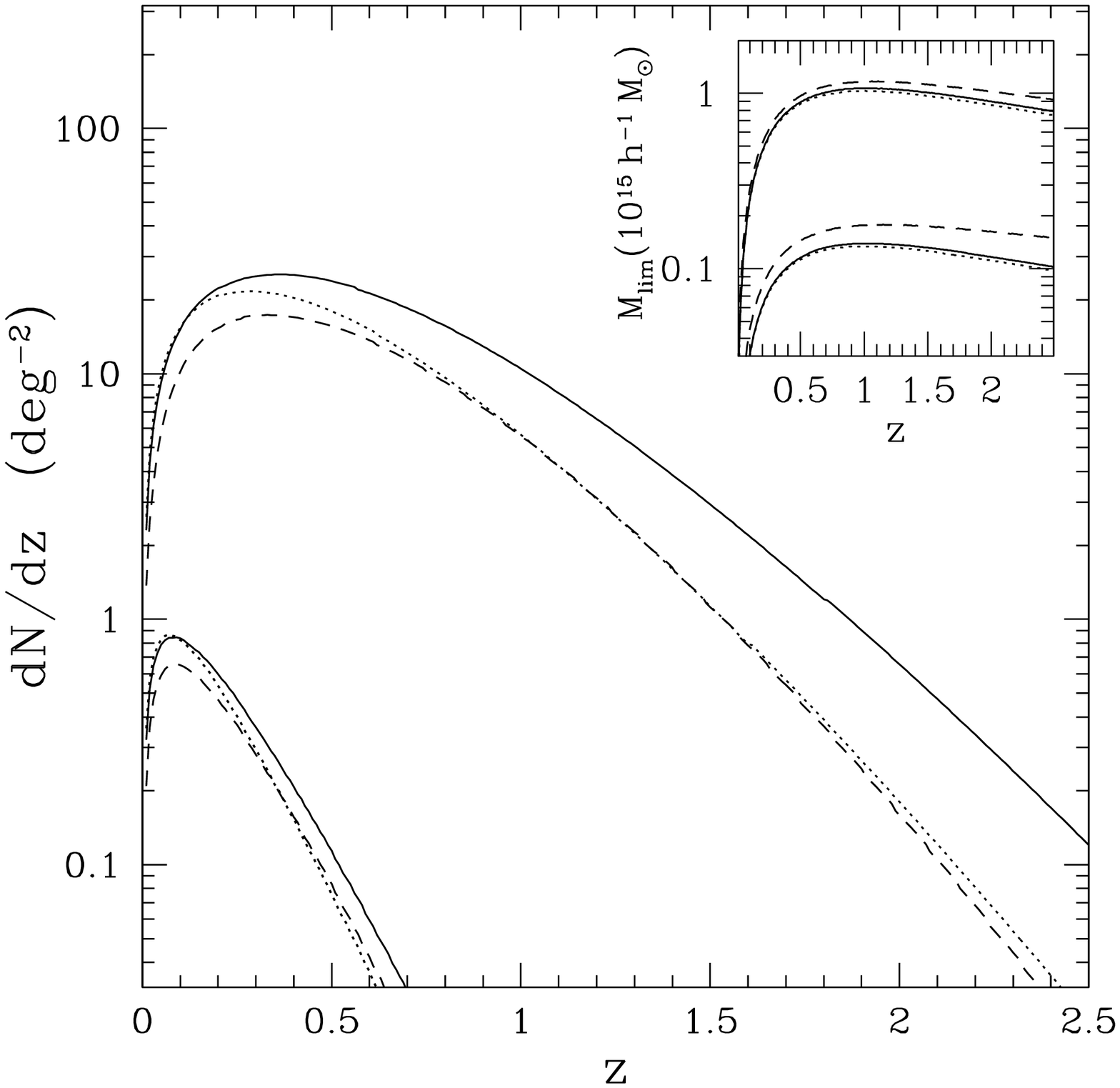}{2.8}{0.45}{-25}
\figcaption{Differential number counts of clusters per square degree and 
mass limits (inset) for SZ surveys for no preheating 
(solid curve) and significant preheating (dashed curve) for a survey 
similar to PLANCK (bottom set of lines for dN/dz and top for mass limits) 
and deep ground-based surveys (top for dN/dz and bottom for mass limits). 
Flux density units are at 30 GHz 
($S \sim 12.7(Y/arcm^2)$). Solid curves assume the $\Lambda CDM$ cosmology
outlined in the text, whereas the dotted curves show self-similar
models with $\Omega_m=0.33$ and $\sigma_8=0.9$ and 0.95 for the top and
bottom lines in the main panel, respectively.
\label{fig:mlims} }

\vskip 0.1in

Redshifts for the clusters will be important to learn about ICM
preheating, but it is interesting that a minimum core radius is imposed
by preheating which is not very sensitive to redshift. The simple existence
of an approximate minimum core radius would be a ``smoking gun'' for preheating,
and the value of this core size is an indicator of the approximate value of
the entropy floor. For example, using $s_{floor}=100$ keV cm$^{2}$
will result in a minimum core radius of roughly 30$''$.
Redshift evolution of the amount of preheating should
be observable in the SZ properties of the cluster catalog. 

Of particular importance is that preheating should not significantly affect the
number of clusters discovered in an SZ survey, only the appearance of the
discovered clusters. However, as indicated by the dotted line in Figure 
\ref{fig:mlims}, the change in expected yields is comparable to the 
change in yields expected from a shift in $\Omega_m$ by 10\%. Clearly,
non-gravitational heating will be an important consideration when trying to
make precise estimates of cosmological parameters from SZ surveys
(e.g., Haiman, Mohr, and Holder 2001).

\section{Small Scale CMB Anisotropy} 
\label{sec:cl}

The structure of high-redshift clusters  can have important implications
for studies of CMB anisotropies. At small angular scales, the
angular power spectrum can become dominated by the thermal SZ effect. 
The magnitude of the thermal SZ power spectrum
is most affected by preheating at exactly the angular scales where it
becomes dominant, as shown in Figure \ref{fig:sz_cl}, as also found in 
other studies \citep{holder99a,komatsu99a,springel00}.

\myputfigure{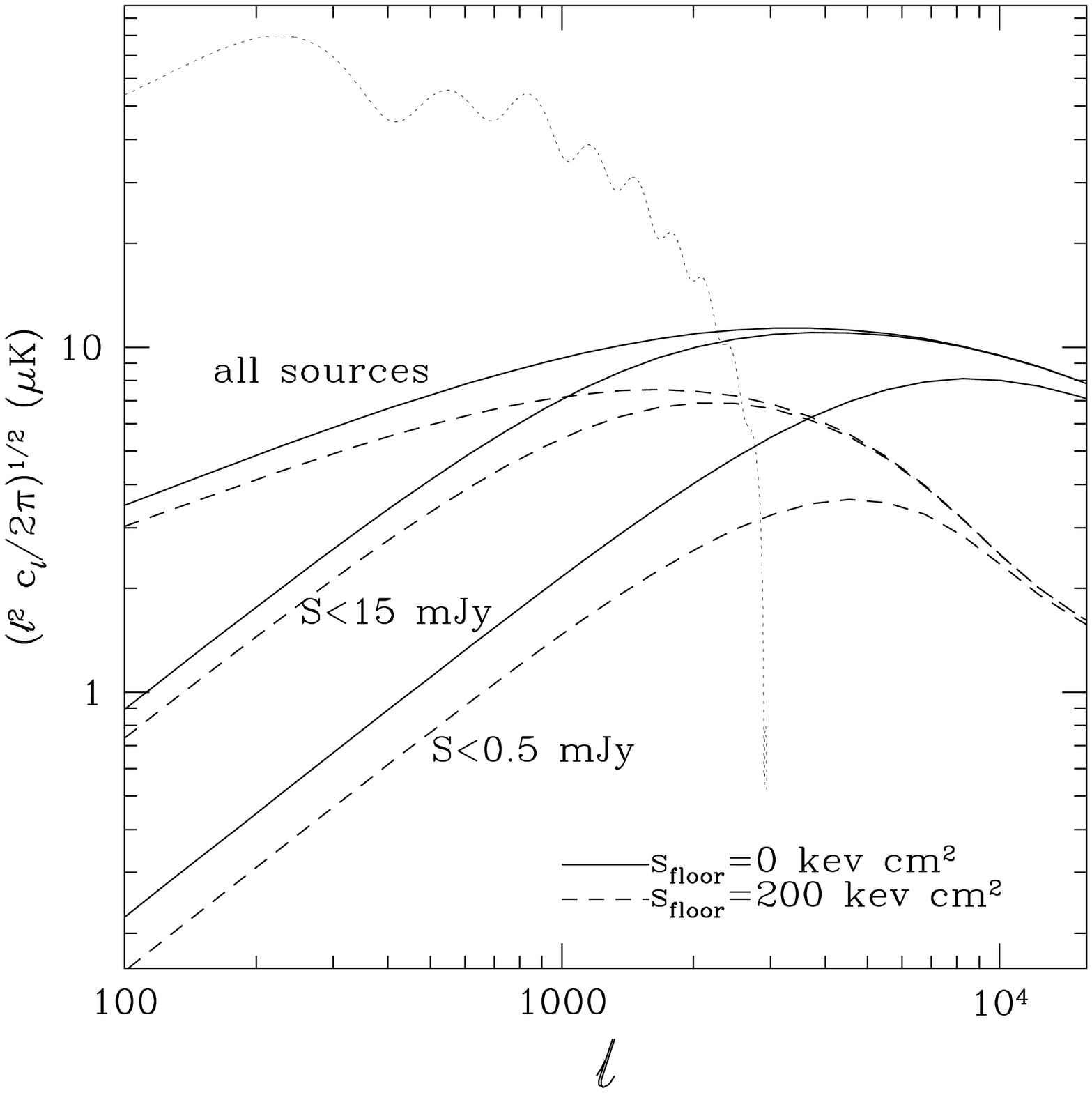}{2.8}{0.45}{-25}
\figcaption{Angular power spectrum from SZ clusters. Top solid curve shows
the power spectrum due to all SZ clusters in the self-similar model,
while the top dashed curve shows the power spectrum from clusters assuming
a model with significant preheating. 
The middle solid curve shows the residual power spectrum for the 
self-similar model after all 
SZ clusters brighter than 15 mJy (at 30 GHz) have been removed, 
with the middle dashed line showing the preheated model.
The bottom solid line and dashed curves correspond to the self-similar 
and preheated models, respectively, after subtraction of all SZ clusters
with flux brighter than 0.5 mJy. 
The expected contribution from primary anisotropies (unlensed) is shown as
the light dotted curve, generated by CMBFast (Seljak and Zaldarriaga 1996).
\label{fig:sz_cl}  }

\vskip 0.1in

Subtracting out the bright clusters can significantly reduce the power
at low multipoles, but does not significantly
affect the angular power spectrum at high $\ell$, 
where the thermal SZ effect is expected to exceed the primary anisotropies.
The power at high $\ell$ is mainly due to low-mass clusters at higher $z$, 
where non-gravitational heating can be very significant. The higher densities
at higher redshift lead to a lower self-similar prediction for the
central entropy, making clusters at higher redshift especially sensitive
to excess entropy.
Deep SZ surveys will allow studies of the clusters which are contributing
the majority of the signal at small angular scales. Shallow surveys will
detect the clusters which provide the bulk of the power at larger angular
scales, where the primary anisotropies are more than an order of magnitude 
larger than the thermal SZ effect, but will have very little information
on the small angular scales at $\ell \ga 2000$.

\section{Discussion and Conclusions}
\label{sec:disc}

We have shown that the SZ effect is well-suited for studies of
non-gravitational heating. A
well-planned survey should not have a selection function which strongly
depends on the amount of non-gravitational heating, while SZ images
of the resulting catalog will be very sensitive to this energy injection.
As a redshift-independent probe, the SZ is particularly well-suited for
high-redshift work, where non-gravitational heating can be very
important. 

We have shown that the expected yield and redshift distribution of clusters
for an SZ survey is only weakly
sensitive to preheating, but that this sensitivity is significant if one wishes
to do ``precision cosmology'' with SZ surveys. This is true for either
deep or shallow surveys. While high-mass clusters are less sensitive to
preheating, the survey yield is more sensitive to the mass limit at higher
masses. 
Thus we have a situation where it will be difficult to use high-mass clusters 
to learn about preheating, but we must understand it to do cosmology. 
For low-mass clusters, the mass limit is still
important for the survey yield, but direct imaging of the resulting catalog
of clusters should yield a wealth of information on preheating.

The peak of the SZ angular power spectrum is very sensitive to the details
of non-gravitational heating and direct images of high redshift, low mass
clusters may be the best way of understanding this signal. Approximately
half of the power at $\ell \ga 2000$, where the thermal SZ signal is stronger
than the primary anisotropies, should be coming from clusters that could
be detected in deep SZ surveys. 

Upcoming CMB experiments such as MAP and PLANCK, with their large
beam sizes, are not well-suited for finding high redshift and low mass clusters.
Ground-based bolometer arrays and interferometers with sub-arcminute
resolution will be well equipped for finding and studying these 
clusters.
Deep SZ surveys will happen within the next few years and the 
information that they yield will be very valuable for constraining
non-gravitational heating through galaxy formation or reionization. 
This information 
should be remarkably unbiased, and improvements in SZ imaging over the
next several years should continue to improve our understanding of these
processes.

\nocite {seljak96}
\acknowledgements{
This is work supported by NASA LTSA grant number NAG5-7986 and the DOE at 
Chicago.
We acknowledge many useful discussions with Joe Mohr and thank U. Seljak
and M. Zaldarriaga for making CMBFast publicly available.
}

\bibliographystyle{apj}


\end{document}